\documentclass[sn-mathphys,Numbered]{sn-jnl}

\usepackage{graphicx}%
\usepackage{multirow}%
\usepackage{amsmath,amssymb,amsfonts}%
\usepackage{amsthm}%
\usepackage{mathrsfs}%
\usepackage[title]{appendix}%
\usepackage{xcolor}%
\usepackage{textcomp}%
\usepackage{manyfoot}%
\usepackage{booktabs}%
\usepackage{algorithm}%
\usepackage{algorithmicx}%
\usepackage{algpseudocode}%
\usepackage{listings}%
\usepackage{subcaption}
\usepackage{caption}

\theoremstyle{thmstyleone}%
\theoremstyle{thmstyletwo}%

\theoremstyle{thmstylethree}%

\begin{document}

\title[Finite Dip-slip Fault in Fractional Burger Rheology]{Determination of Effect of the Movement of a Finite, Dip-slip Fault in Viscoelastic Half-space of Fractional Burger Rheology}

\author*[1]{\fnm{Pabita} \sur{Mahato}}\email{pabitamahato012@gmail.com,\\ https://orcid.org/0000-0002-6859-2140}

\author[2]{\fnm{Seema} \sur{Sarkar (Mondal)}}\email{ssarkarmondal.maths@nitdgp.ac.in}
\equalcont{This author contributed equally to this work.}
\affil[1,2]{\orgdiv{Department of Mathematics}, \orgname{National Institute of Technology Durgapur}, \orgaddress{\city{Durgapur}, \postcode{713209}, \state{West Bengal}, \country{India}}}

\abstract{The seismically active regions often correlate with fault lines, and the movement of these faults plays a crucial role in defining how stress is stored or released in these areas. To investigate the deformation and accumulation/release of stress and strain in seismically active regions during the aseismic period, a mathematical model has been developed by considering a finite, creeping dip-slip fault inclined in the viscoelastic half-space of a fractional Burger rheology. Laplace transformation for fractional derivatives, Modified Green’s function technique, correspondence principle and finally, inverse Laplace transformation have been used to derive analytical solutions for displacement, stress and strain components. The graphical representations were depicted using MATLAB to understand the effect on displacement, stresses and strains due to changes in inclinations and creep velocities of the fault, as well as orders of the fractional derivative. 
Our investigation indicates that a change in creep velocity and inclination of the fault has a significant effect, while a change in the order of fractional derivative has a moderate effect on displacement, stress, and strain components. Analysis of these results can provide insights into subsurface deformation and its impact on fault movement, which can lead to earthquakes.}



\maketitle

\section{Introduction}\label{sec1}
There is increasing interest in examining the impact of static or quasistatic displacements, stresses, and strains on earthquake processes. Seismologists and geological engineers have particularly focused on modelling the dynamic processes triggered by earthquakes. Observations indicate that in seismically active zones, two successive seismic events are typically separated by a long aseismic interval. During this period, advanced instruments like strain and tilt meters detect slow and steady aseismic surface fluctuations. These fluctuations indicate gradual changes in stress and strain in the region, which may eventually lead to sudden or creeping movements along seismic faults. Hence, these faults, whether strike-slip or dip-slip, finite or infinite, surface-breaking or buried, require vivid investigation. To understand earthquake mechanisms, it is essential to develop mathematical models that study the minor ground deformations observed during the aseismic periods in seismically active regions.\\

The foundational work on static ground deformation in elastic media was initiated by Steketee (1958), Chinnery (1961, 1972), and Maruyama (1964). The vertical displacement of the free surface of dip-slip faults has been studied by Savage (1966). Later on, Sen and Debnath (2012) examined long dip-slip faults within a viscoelastic medium characterized by Maxwell-type materials. Mondal et al. (2018) and Kundu et al. (2021) extended this study by studying infinite and finite dip-slip faults in standard linear solid and Burger Rheology's material. Recently, Mondal and Debsarma (2023)  investigated the impact on the stress-strain and displacement components in the localized area of the fault plane by taking an infinite strike-slip fault. They have used classical integer derivatives to represent the mathematical model in all such works.\\

However, in the last few decades, it has been observed that viscoelasticity is the most extensive application of fractional differential and fractional integral operators (Mainardi 1997; Rossikhin and Shitikova 1997; Podlubny 1999; Koeller 1984). Studies show that the integer derivative of a function is influenced only by its nearby points, while the fractional derivative considers non-local features, such as 'infinite memory' (Cafagna 2007; Pelap et al. 2018). Mahato et al. (2022), Mahato and Sarkar Mondal (2025) explored the impact of fractional derivatives on the displacement, stress, and strain of different types of faults located in a viscoelastic half-space of a standard linear solid medium.\\

Fractional derivetives captures the memory and hereditary properties of different materials and processes, which are often neglected by classical integer-order models. Incorporating fractional derivatives into viscoelastic constitutive equations enhances the accuracy of modelling complex time-dependent stress-strain relationships (Deng and Morozov, 2018). Due to their non-local characteristics, fractional derivatives are particularly well-suited for capturing the intricate dynamics of ruptures, such as those occurring during earthquakes (Wu et al., 2023; Fogang et al., 2021). This consideration suggests the incorporation of fractional differential equations for modelling problems in viscoelastic systems. In this paper, we have considered a single finite buried fault situated in the viscoelastic half-space of fractional Burger Rheology. The fault is an inclined dip-slip fault following creeping movement. The paper is organized as follows: In sections \ref{sec2}, the formulation of the theoretical model has been explained. The analytical solution of the boundary value problem of the model is described in section \ref{sec3}. Section \ref{sec4} exhibits the results of our numerical search through MATLAB R2022b. Section \ref{sec5} deals with the concluding remarks and future scope of the research, followed by a declaration and reference list.
\section{Formulation}\label{sec2}
\begin{figure}[ht]
\begin{subfigure}{.5\textwidth}
  \centering
  \includegraphics[width=1\linewidth]{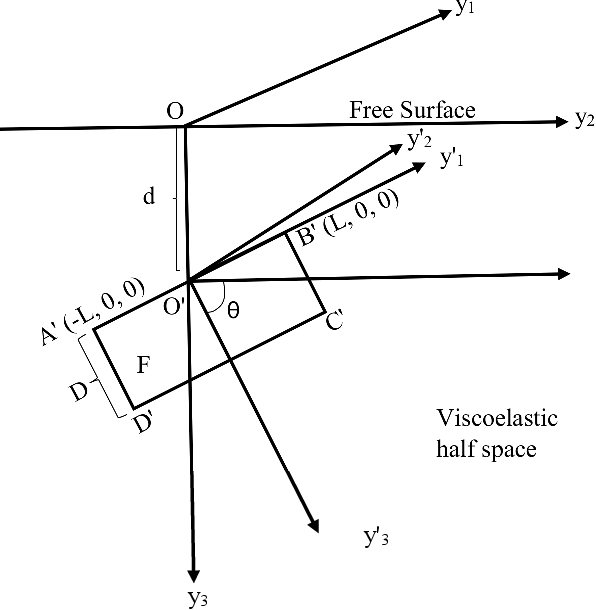}
  \caption{}
  \label{figure1a}
\end{subfigure}%
\begin{subfigure}{.5\textwidth}
  \centering
  \includegraphics[width=1\linewidth]{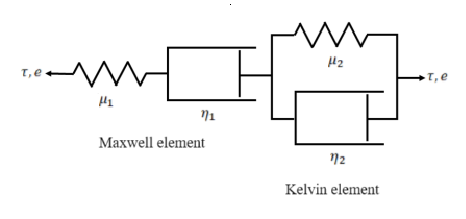}
  \caption{}
  \label{figure1b}
\end{subfigure}
\caption{(a) Semantic diagram of finite dip-slip fault in fractional Burger rheology. (b) Spring-dashpot representation of the fractional Burger’s rheology.}
\label{figure1}
\end{figure}

We have considered a three-dimensional model taking a finite, inclined, buried, dip-slip fault situated in the viscoelastic half-space of fractional Burger rheology. Let $A'B'C'D'$ be the fault $F$ of length $2L$ ($L$ is finite), which is located at a depth $d$ below the free surface. Also, we assume that $D$ is the width of the fault, and it is inclined at an angle $\theta$ with the free surface, as shown in figure \ref{figure1}.

To understand the position of the fault, we have considered a Cartesian coordinate system $(y_1, y_2, y_3)$ with origin $O$ so that $y_3=0$ represents the free surface, $y_1$ axis chosen along a straight line vertically above the upper edge of the fault and $y_2$ axis is perpendicular to $y_1$ axis lying on the plane $y_3=0$. Thus $y_3\geq 0$ represents the viscoelastic half-space. As the upper edge of the fault is at a depth $d$ below the free surface and the fault is also inclined at a certain angle $\theta$ with the free surface, we have considered one more coordinate system $(y_1', y_2', y_3')$ with origin $O'$. Here $y_1'$ axis is parallel to $y_1$ i.e., along the length of the fault, $y_3'$ axis is along the width of the fault and $y_2'$ axis is perpendicular to the plane $y_1'y_3'$. The coordinates of $A'$ and $B'$ with respect to the new coordinate system are $A'(-L, 0,0)$ and $B'(L, 0, 0)$ respectively. Hence, the position of the fault $F$ is defined by $F: -L\leq y_1'\leq L, y_2'=0, 0\leq y_3'\leq D$ and the relation between the two coordinate systems is given by
\begin{eqnarray}
\begin{cases}
    y_1'=y_1\\
    y_2'=y_2\sin\theta-(y_3-d)\cos\theta\\
    y_3'=y_2\cos\theta+(y_3-d)\sin\theta
    \end{cases}
    \label{equation1}
\end{eqnarray}
Let $u_i$ be the displacement components along $y_i$, $\tau_{ij}$ be the stress components produced by an internal force in the direction of $y_j$, acting on a surface, having a normal in the direction of $y_i$ and $e_{ij}$ be the associated strain components, where $i, j=1, 2, 3$. The displacement, stress and strain components associated with finite dip-slip fault are $(u_1, u_2, u_3, \tau_{11}, \tau_{12}, \tau_{13}, \tau_{22}, \tau_{23}, \tau_{33}, e_{11}, e_{12}, e_{13}, e_{22}, e_{23}, e_{33})$ as explained by Kundu et al. (2021). All of them are functions of $y_1, y_2, y_3$ and time $t$.

Our observation begins at a moment when there is no fault movement within the medium, and the entire viscoelastic medium is in a quasi-static, aseismic state. However, a slow subsurface deformation is continuous within it. The medium experiences the impact of shear stresses resulting from dip-slip fault movement, which are sustained by various tectonic forces, including mantle convection, the weight of overlying materials, internal pressure within the Earth, and other geological transformations. Then, the displacement, stress and strain components associated with finite dip-slip fault satisfy some constitutive equations, stress equation of motion and boundary conditions for $t\geq 0$, which are depicted below. 
\subsection{Constitutive Equation}\label{sec2.1}
The constitutive equation, which describes the relationship between stress and strain, including the time derivative, for a finite dip-slip fault located in a viscoelastic half-space with Burger rheology, is presented below (suggested by Okuka and Zorica 2020; Segall 2010).
\begin{eqnarray}
  \tau_{ij}+p_1\, {_0D^{\alpha}_t} (\tau_{ij})+p_2\,{_0D^{2\alpha}_t} (\tau_{ij})=2q_1\,{_0D^{\alpha}_t} (e_{ij})+2q_2\, _0D^{2\alpha}_t (e_{ij}),\, i, j=1, 2, 3
    \label{equation2}
\end{eqnarray}
where $p_1=\frac{\eta_1}{\mu_1}+\frac{\eta_2}{\mu_2}+\frac{\eta_1}{\mu_2}$, $p_2=\frac{\eta_1\eta_2}{\mu_1\mu_2}$, $q_1=\eta_1$, $q_2=\frac{\eta_1\eta_2}{\mu_2}$ such that $\eta_1, \eta_2$ are effective viscosities and $\mu_1, \mu_2$ are effective rigidities of the materials and are constant throughout the entire medium. The strain components $e_{ij}$ is defined by 
$e_{ij}=\frac12\left(\frac{\partial u_i}{\partial y_j}+\frac{\partial u_j}{\partial y_i}\right)$. $_0D^{\alpha}_t$ is fractional operator of order $\alpha$ ($0<\alpha\leq1$) and is defined by\\
$\left(_0D^{\alpha}y\right)(x)=\frac1{\Gamma(n-\alpha)}\int_0^t \frac{y^{(n)}(t)}{(x-t)^{\alpha-n+1}}dt$ (Caputo, 1969)\\
such that
    $n=\begin{cases}
        [Re(\alpha)]+1 \mbox{ for }\alpha\notin \mathbb N_0\\ \alpha \mbox{ for } \alpha \in \mathbb N_0
    \end{cases}$\\
$\mathbb N_0$ is the set of natural numbers including $0$.
\subsection{Stress equation of motion}\label{sec2.2}
Initially, at $t=0$, no fault movement is assumed to occur within the medium, but a slow, aseismic quasi-static deformation occurs over time. Due to this deformation, the effect of inertial forces is deemed negligible. Also, no change of body forces is supposed to occur over the medium with respect to the initial stress level, and therefore, the stress equation of motion related to finite dip-slip fault becomes
\begin{eqnarray}
\begin{cases}
    \frac{\partial \tau_{11}}{\partial y_1}+\frac{\partial \tau_{12}}{\partial y_2}+\frac{\partial \tau_{13}}{\partial y_3}=0\\
    \frac{\partial \tau_{21}}{\partial y_1}+\frac{\partial \tau_{22}}{\partial y_2}+\frac{\partial \tau_{23}}{\partial y_3}=0\\
    \frac{\partial \tau_{31}}{\partial y_1}+\frac{\partial \tau_{32}}{\partial y_2}+\frac{\partial \tau_{33}}{\partial y_3}=0
\end{cases}
    \label{equation3}
\end{eqnarray}
where $-L\leq y_1\leq L, -\infty<y_2<\infty$ and $y_3\geq0$.
\subsection{Boundary conditions}\label{sec2.3}
The boundary conditions for the dip-slip fault are taken as follows:\\
At the tip of the fault, i.e., at the point $A'$ and $B'$, the limiting value of normal stress $\tau_{11}$ gives a constant value.
\begin{eqnarray}
    \lim_{y_1'\to L^-}\tau_{11}=\lim_{y_1'\to L^+}\tau_{11}=\tau_{L}, \, y_2'=0, 0\leq y_3'\leq D, t\geq 0
    \label{equation4}
    \end{eqnarray}
    \begin{eqnarray}
    \lim_{y_1'\to {-L}^-}\tau_{11}=\lim_{y_1'\to -L^+}\tau_{11}=\tau_{-L}, \, y_2'=0, 0\leq y_3'\leq D, t\geq 0
     \label{equation5}
\end{eqnarray}
Here, we consider that stress maintains constant values at the tip of the fault along the $y_1'$ axis. These constants are likely to be small enough so that no further extension is possible along the $y_1'$-axis, i.e., $\tau_{L}=\tau_{-L}$.\\
Stress components $\tau_{22}, \tau_{33}$ increase very slowly with time. i.e.,
\begin{eqnarray}
    \lim_{|y_2|\to\infty}\tau_{22}=\tau_{\infty}(0)(1+kt)\cos\theta, k>0 \mbox{ for}-L\leq y_1\leq L, y_3\geq0, t\geq0
    \label{equation6}
\end{eqnarray}
and
\begin{eqnarray}
     \lim_{|y_2|\to\infty}\tau_{33}=\tau_{\infty}(0)(1+kt)\sin\theta, k>0 \mbox{ for}-L\leq y_1\leq L, y_3\geq0, t\geq0
    \label{equation7}
\end{eqnarray}
where $k$ is constant and is very small.\\
No stress is transferred between the solid earth and the atmosphere. So other stress components vanish on the earth's surface, i.e.,
\begin{eqnarray}
   \tau_{12}=0, \tau_{13}=0, \tau_{23}=0 \mbox{ for } -L\leq y_1\leq L, -\infty\leq y_2\leq\infty, t\geq0
    \label{equation8}
\end{eqnarray}
Also, these stress components $(\tau_{12}, \tau_{13}, \tau_{23})$ tend to zero as we move vertically downwards below the free surface away from the fault.
i.e.,\\
As $y_3\to\infty$,
\begin{eqnarray}
   \tau_{12}\to0, \tau_{13}\to0, \tau_{23}\to0 \mbox{ for } -L\leq y_1\leq L, -\infty\leq y_2\leq\infty, t\geq0
    \label{equation9}
    \end{eqnarray}    
\subsection{Initial conditions}\label{sec2.4}
As mentioned earlier, we start our observation at an instant $t=0$. At that instant, displacement, stress and strain components are taken as ${(u_i)}_0, {(\tau_{ij})}_0, {(e_{ij})}_0$ respectively $(i, j=1, 2, 3)$. All of them are either constants or functions of $y_1, y_2, y_3$ and satisfy all the above relations from (\ref{equation2}) to (\ref{equation9}).
\section{Solutions}\label{sec3} 
For $i=1$ and $j=1, 2, 3$, equation (\ref{equation2}) is of the form
\begin{eqnarray}
    \begin{cases}       
  \tau_{11}+p_1\, {_0D^{\alpha}_t} (\tau_{11})+p_2\,{_0D^{2\alpha}_t} (\tau_{11})=2q_1\,{_0D^{\alpha}_t} (e_{11})+2q_2\, _0D^{2\alpha}_t (e_{11})\\
  \tau_{12}+p_1\, {_0D^{\alpha}_t} (\tau_{12})+p_2\,{_0D^{2\alpha}_t} (\tau_{12})=2q_1\,{_0D^{\alpha}_t} (e_{12})+2q_2\, _0D^{2\alpha}_t (e_{12})\\
  \tau_{13}+p_1\, {_0D^{\alpha}_t} (\tau_{13})+p_2\,{_0D^{2\alpha}_t} (\tau_{13})=2q_1\,{_0D^{\alpha}_t} (e_{13})+2q_2\, _0D^{2\alpha}_t (e_{13})
    \end{cases}
    \label{equation10}
\end{eqnarray}
which also can be expanded as
\begin{eqnarray}
    \begin{cases}       
  \tau_{11}+p_1\, {_0D^{\alpha}_t} (\tau_{11})+p_2\,{_0D^{2\alpha}_t} (\tau_{11})=2q_1\,{_0D^{\alpha}_t} \left(\frac{\partial u_1}{\partial y_1}\right)+2q_2\, _0D^{2\alpha}_t \left(\frac{\partial u_1}{\partial y_1}\right)\\
  \tau_{12}+p_1\, {_0D^{\alpha}_t} (\tau_{12})+p_2\,{_0D^{2\alpha}_t} (\tau_{12})=2q_1\,{_0D^{\alpha}_t} \left(\frac{\partial u_1}{\partial y_2}\right)+2q_2\, _0D^{2\alpha}_t \left(\frac{\partial u_1}{\partial y_2}\right)\\
  \tau_{12}+p_1\, {_0D^{\alpha}_t} (\tau_{12})+p_2\,{_0D^{2\alpha}_t} (\tau_{12})=2q_1\,{_0D^{\alpha}_t} \left(\frac{\partial u_2}{\partial y_1}\right)+2q_2\, _0D^{2\alpha}_t \left(\frac{\partial u_2}{\partial y_1}\right)\\
  \tau_{13}+p_1\, {_0D^{\alpha}_t} (\tau_{13})+p_2\,{_0D^{2\alpha}_t} (\tau_{13})=2q_1\,{_0D^{\alpha}_t} \left(\frac{\partial u_1}{\partial y_3}\right)+2q_2\, _0D^{2\alpha}_t \left(\frac{\partial u_1}{\partial y_3}\right)\\
  \tau_{13}+p_1\, {_0D^{\alpha}_t} (\tau_{13})+p_2\,{_0D^{2\alpha}_t} (\tau_{13})=2q_1\,{_0D^{\alpha}_t} \left(\frac{\partial u_3}{\partial y_1}\right)+2q_2\, _0D^{2\alpha}_t \left(\frac{\partial u_3}{\partial y_1}\right)
    \end{cases}
    \label{equation11}
\end{eqnarray}
 Now, for deriving the analytical solution from these governing equations, we differentiate the 1st, 2nd and 4th relation of constitutive equations (\ref{equation11}) partially with respect to $y_1$, $y_2$ and $y_3$ respectively and add them. Then, using the relation (\ref{equation3}) and the initial conditions, we get
 \begin{eqnarray}
     \nabla^2 U_1=0, \mbox{ where } U_1=u_1-{(u_1)}_0
     \label{equation12}
 \end{eqnarray}
 Similarly, using the other constitutive equations i.e., for $i=2, 3$ and $j=1, 2, 3$ of (\ref{equation2}), we get
 \begin{eqnarray}
     \nabla^2 U_2=0, \mbox{ where } U_2=u_2-{(u_2)}_0
     \label{equation13}
 \end{eqnarray}
  \begin{eqnarray}
     \nabla^2 U_3=0, \mbox{ where } U_3=u_3-{(u_3)}_0
     \label{equation14}
 \end{eqnarray}
 We will now solve the above time-dependent boundary value problem before and after the fault movement with the help of Laplace transformation.
 
The final solution for displacement, stress and strain components will be represented in the form
\begin{eqnarray}
\begin{cases}
    u_i={(u_i)}_1+{(u_i)}_2,\, i=1, 2, 3\\
    \tau_{ij}={(\tau_{ij})}_1+{(\tau_{ij})}_2,\, i, j=1, 2, 3\\
    e_{ij}={(e_{ij})}_1+{(e_{ij})}_2,\, i, j=1, 2, 3
    \end{cases}
    \label{equation15}
\end{eqnarray}
where ${(u_i)}_1, {(\tau_{ij})}_1, {(e_{ij})}_1$, $(i, j = 1, 2, 3)$ are displacement, stress and strain components in the absence of the movement of the fault and are all continuous throughout the system while ${(u_i)}_2, {(\tau_{ij})}_2, {(e_{ij})}_2$, $(i, j = 1, 2, 3)$ are those of after the movement of the fault.
 \subsection{Solution before the movement of the fault}\label{sec3.1}
The components ${(u_i)}_1, {(\tau_{ij})}_1, {(e_{ij})}_1$, $(i, j = 1, 2, 3)$ can be found by solving boundary value problem (\ref{equation2})-(\ref{equation9}), (\ref{equation12})-(\ref{equation14}) with the help of Laplace transformation.\\ 
First taking Laplace transformation with respect to time $t$ on the equations (\ref{equation12})-(\ref{equation14}) we get
\begin{eqnarray}
    \begin{cases}
         \nabla^2 \overline {U_1}=0, \mbox{ where } \overline{U}_1=\overline{u_1}-\frac{{(u_1)}_0}s\\
         \nabla^2 \overline {U_2}=0, \mbox{ where } \overline{U_2}=\overline{u_2}-\frac{{(u_2)}_0}s\\
         \nabla^2 \overline {U_3}=0, \mbox{ where } \overline{U_3}=\overline{u_3}-\frac{{(u_3)}_0}s
    \end{cases}
    \label{equation16}
\end{eqnarray}
where $s$ is the Laplace transform variable to time $t$ and $\overline{u_i}$, $\overline{U_i}$ $(i=1, 2, 3)$ are Laplace transforms of $u_i$, $U_i$ respectively.\\    
Let the trial solutions of (\ref{equation16}) be
\begin{eqnarray}
\begin{cases}
\overline{u_1}=\frac{{(u_1)}_0}s+A_1y_1+B_1y_2+C_1y_3\\
    \overline{u_2}=\frac{{(u_2)}_0}s+A_2y_1+B_2y_2+C_2y_3\\
    \overline{u_3}=\frac{{(u_3)}_0}s+A_3y_1+B_3y_2+C_3y_3
    \label{equation17}
\end{cases}    
    \end{eqnarray}
where $A_i, B_i, C_i, i=1, 2, 3$ are constants or function of $s$ in transformed domain.\\ 
Taking Laplace transformation on all the constitutive equations (\ref{equation2}) for $i,j=1,2,3$ (neglecting first and higher order derivatives of $\tau_{ij}$) and using (\ref{equation17}) we get 
\begin{eqnarray}
    \begin{cases}
    \overline{\tau_{11}}(1+p_1s^{\alpha}+p_2s^{2{\alpha}})-{(\tau_{11})}_0(p_1s^{\alpha-1}+p_2s^{2\alpha-1})=A_1(2q_1s^{\alpha}+2q_2s^{2\alpha})\\
    \overline{\tau_{12}}(1+p_1s^{\alpha}+p_2s^{2{\alpha}})-{(\tau_{12})}_0(p_1s^{\alpha-1}+p_2s^{2\alpha-1})=(B_1+A_2)(q_1s^{\alpha}+q_2s^{2\alpha})\\
    \overline{\tau_{13}}(1+p_1s^{\alpha}+p_2s^{2{\alpha}})-{(\tau_{13})}_0(p_1s^{\alpha-1}+p_2s^{2\alpha-1})=(C_1+A_3)(q_1s^{\alpha}+q_2s^{2\alpha})\\
    \overline{\tau_{22}}(1+p_1s^{\alpha}+p_2s^{2{\alpha}})-{(\tau_{22})}_0(p_1s^{\alpha-1}+p_2s^{2\alpha-1})=B_2(2q_1s^{\alpha}+2q_2s^{2\alpha})\\
    \overline{\tau_{23}}(1+p_1s^{\alpha}+p_2s^{2{\alpha}})-{(\tau_{23})}_0(p_1s^{\alpha-1}+p_2s^{2\alpha-1})=(C_2+B_3)(q_1s^{\alpha}+q_2s^{2\alpha})\\
    \overline{\tau_{33}}(1+p_1s^{\alpha}+p_2s^{2{\alpha}})-{(\tau_{33})}_0(p_1s^{\alpha-1}+p_2s^{2\alpha-1})=C_3(2q_1s^{\alpha}+2q_2s^{2\alpha})
    \end{cases}
    \label{equation18}
\end{eqnarray}
where $\overline{\tau_{ij}}$ $(i,j=1, 2, 3)$ are Laplace transforms of $\tau_{ij}$.\\
Now, taking the Laplace transformation on all the boundary conditions (\ref{equation4})-(\ref{equation9}), we get
\begin{eqnarray}
    \lim_{y_1'\to L^-}\overline{\tau_{11}}=\lim_{y_1'\to L^+}\overline{\tau_{11}}=\frac{\tau_{L}}s, \, y_2'=0, 0\leq y_3'\leq D, t\geq 0
    \label{equation19}
    \end{eqnarray}
    \begin{eqnarray}
    \lim_{y_1'\to {-L}^-}\overline{\tau_{11}}=\lim_{y_1'\to -L^+}\overline{\tau_{11}}=\frac{\tau_{L}}s, \, y_2'=0, 0\leq y_3'\leq D, t\geq 0
     \label{equation20}
\end{eqnarray}
\begin{eqnarray}
\lim_{|y_2|\to\infty}\overline{\tau_{22}}=\tau_{\infty}(0)\left(\frac1s+\frac k{s^2}\right)\cos\theta, k>0 \mbox{ for}-L\leq y_1\leq L, y_3\geq0, t\geq0
    \label{equation21}
\end{eqnarray}
\begin{eqnarray}
\lim_{|y_2|\to\infty}\overline{\tau_{33}}=\tau_{\infty}(0)\left(\frac1s+\frac k{s^2}\right)\sin\theta, k>0 \mbox{ for}-L\leq y_1\leq L, y_3\geq0, t\geq0
    \label{equation22}
\end{eqnarray}
$k$ is constant and is very small.
\begin{eqnarray}
   \overline{\tau_{12}}=0, \overline{\tau_{13}}=0, \overline{\tau_{23}}=0 \mbox{ for } -L\leq y_1\leq L, -\infty\leq y_2\leq\infty, t\geq0
    \label{equation23}
\end{eqnarray} 
\begin{eqnarray}
   \overline{\tau_{12}}\to0, \overline{\tau_{13}}\to0, \overline{\tau_{23}}\to0 \mbox{ for } -L\leq y\leq L, -\infty\leq y_2\leq\infty, t\geq0
    \label{equation24}
    \end{eqnarray}
Hence, substituting all the above-transformed boundary conditions in equation (\ref{equation18}), we get the unknowns as
\begin{eqnarray}
    \begin{cases}
        A_1=\frac{{\tau}_L}{s(2q_1s^{\alpha}+2q_2s^{2\alpha})}\\
        B_1=0\\
        C_1=0\\
        A_2=0\\
        B_2=\frac{\tau_{\infty}(0)\cos\theta}{2q_1s^{\alpha}+2q_2s^{2\alpha}}\left(\frac1s+\frac{k}{s^2}+p_1ks^{\alpha-2}+p_2ks^{2\alpha-2}\right)\\
        C_2=0\\
        A_3=0\\
        B_3=0\\
        C_3=\frac{\tau_{\infty}(0)\sin\theta}{2q_1s^{\alpha}+2q_2s^{2\alpha}}\left(\frac1s+\frac{k}{s^2}+p_1ks^{\alpha-2}+p_2ks^{2\alpha-2}\right)
    \end{cases}
    \label{equation25}
\end{eqnarray}
Now, substituting the expressions of $A_i, B_i, C_i$ $(i=1, 2, 3)$ in (\ref{equation17}) and (\ref{equation18}) and taking inverse Laplace transformation we get the displacement and stress components before any fault movement as
\begin{eqnarray}
    \begin{cases}
        {(u_1)}_1={(u_1)}_0+\frac{\tau_Ly_1}{2q_1}\left[\frac{t^{\alpha}}{\Gamma(\alpha+1)}-\frac{q_2}{q_1}\left(1-E_{\alpha}\left(-\frac{q_1}{q_2}t^{\alpha}\right)\right)\right]\\
        {(u_2)}_1={(u_2)}_0+\frac{\tau_{\infty}(0)y_2\cos\theta}{2q_1}\Big[\frac{t^{\alpha+1}}{\Gamma(\alpha+2)}+\frac{t^{\alpha}}{\Gamma(\alpha+1)}+kt\left(p_1-\frac{q_2}{q_1}\right)\cr\hspace{5mm}+\left\{E_{\alpha }\left(-\frac{q_1}{q_2}t^{\alpha}\right)-1\right\}\left\{\frac{q_2}{q_1}+\frac{kq_2}{q_1}\left(p_1-\frac{q_1p_2}{q_2}-\frac{q_2}{q_1}\right)\right\}\Big]\\
        {(u_3)}_1={(u_3)}_0+\frac{\tau_{\infty}(0)y_3\sin\theta}{2q_1}\Big[\frac{t^{\alpha+1}}{\Gamma(\alpha+2)}+\frac{t^{\alpha}}{\Gamma(\alpha+1)}+kt\left(p_1-\frac{q_2}{q_1}\right)\cr\hspace{5mm}+\left\{E_{\alpha }\left(-\frac{q_1}{q_2}t^{\alpha}\right)-1\right\}\left\{\frac{q_2}{q_1}+\frac{kq_2}{q_1}\left(p_1-\frac{q_1p_2}{q_2}-\frac{q_2}{q_1}\right)\right\}\Big]
    \end{cases}
    \label{equation26}
\end{eqnarray}
\begin{eqnarray}
    \begin{cases}
        {(\tau_{11})}_1={(\tau_{11})}_0+\left[1+\frac{1}{\sqrt{p_1^2-4p_2}}\left\{\frac{E_{\alpha}\left(-\lambda_1t^{\alpha}\right)}{-\lambda_1}+\frac{E_{\alpha}\left(-\lambda_2t^{\alpha}\right)}{\lambda_2}\right\}\right]\left\{\tau_L-{(\tau_{11})}_0\right\}\\
       {(\tau_{12})}_1=\frac{{(\tau_{12})}_0}{\sqrt{p_1^2-4p_2}}\left[\frac{E_{\alpha}\left(-\lambda_1t^{\alpha}\right)}{\lambda_1}-\frac{E_{\alpha}\left(-\lambda_2t^{\alpha}\right)}{\lambda_2}\right]\\
         {(\tau_{13})}_1=\frac{{(\tau_{13})}_0}{{\sqrt{p_1^2-4p_2}}}\left[\frac{E_{\alpha}\left(-\lambda_1t^{\alpha}\right)}{\lambda_1}-\frac{E_{\alpha}\left(-\lambda_2t^{\alpha}\right)}{\lambda_2}\right]\\
         {(\tau_{22})}_1=\tau_{\infty}(0)(1+kt)\cos\theta+\frac{1}{\sqrt{p_1^2-4p_2}}\left[\frac{E_{\alpha}\left(-\lambda_1t^{\alpha}\right)}{\lambda_1}-\frac{E_{\alpha}\left(-\lambda_2t^{\alpha}\right)}{\lambda_2}\right]\\\hspace{8mm}\left\{{(\tau_{22})}_0-{\tau_{\infty}}(0)\cos\theta\right\}\\
         {(\tau_{23})}_1=\frac{{(\tau_{23})}_0}{{\sqrt{p_1^2-4p_2}}}\left[\frac{E_{\alpha}\left(-\lambda_1t^{\alpha}\right)}{\lambda_1}-\frac{E_{\alpha}\left(-\lambda_2t^{\alpha}\right)}{\lambda_2}\right]\\
          {(\tau_{33})}_1=\tau_{\infty}(0)(1+kt)\sin\theta+\frac{1}{\sqrt{p_1^2-4p_2}}\left[\frac{E_{\alpha}\left(-\lambda_1t^{\alpha}\right)}{\lambda_1}-\frac{E_{\alpha}\left(-\lambda_2t^{\alpha}\right)}{\lambda_2}\right]\\\hspace{8mm}\left\{{(\tau_{33})}_0-{\tau_{\infty}}(0)\sin\theta\right\}
    \end{cases}
    \label{equation27}
\end{eqnarray}
where $\lambda_1=\frac{p_1-\sqrt{p_1^2-4p_2}}{2p_2}$ and $\lambda_2=\frac{p_1+\sqrt{p_1^2-4p_2}}{2p_2}$, $E_{\alpha}(x)$ is the Mittag-Lefller function (Mittag 1903) defined as $E_{\alpha}(x)=\sum_{i=0}^{\infty}\frac{x^i}{\Gamma(\alpha i+1)}$.\\
Using the relation $e_{ij}=\frac12\left(\frac{\partial u_i}{\partial y_j}+\frac{\partial u_j}{\partial y_i}\right)$ we get the strain components as
\begin{eqnarray}
    \begin{cases}
        {(e_{11})}_1={(e_{11})}_0+\frac{\tau_L}{2q_1}\left[\frac{t^{\alpha}}{\Gamma(\alpha+1)}-\frac{q_2}{q_1}\left(1-E_{\alpha}\left(-\frac{q_1}{q_2}t^{\alpha}\right)\right)\right]\\
        {(e_{12})}_1={(e_{12})}_0,\,
        {(e_{13})}_1={(e_{13})}_0, \, {(e_{23})}_1={(e_{23})}_0\\
        {(e_{22})}_1={(e_{22})}_0+\frac{\tau_{\infty}(0)\cos\theta}{2q_1}\Big[\frac{t^{\alpha+1}}{\Gamma(\alpha+2)}+\frac{t^{\alpha}}{\Gamma(\alpha+1)}+kt\left(p_1-\frac{q_2}{q_1}\right)\cr\hspace{5mm}+\left\{E_{\alpha }\left(-\frac{q_1}{q_2}t^{\alpha}\right)-1\right\}\left\{\frac{q_2}{q_1}+\frac{kq_2}{q_1}\left(p_1-\frac{q_1p_2}{q_2}-\frac{q_2}{q_1}\right)\right\}\Big]\\
        {(e_{33})}_1={(e_{33})}_0+\frac{\tau_{\infty}(0)\sin\theta}{2q_1}\Big[\frac{t^{\alpha+1}}{\Gamma(\alpha+2)}+\frac{t^{\alpha}}{\Gamma(\alpha+1)}+kt\left(p_1-\frac{q_2}{q_1}\right)\cr\hspace{5mm}+\left\{E_{\alpha }\left(-\frac{q_1}{q_2}t^{\alpha}\right)-1\right\}\left\{\frac{q_2}{q_1}+\frac{kq_2}{q_1}\left(p_1-\frac{q_1p_2}{q_2}-\frac{q_2}{q_1}\right)\right\}\Big]
    \end{cases}
    \label{equation28}
\end{eqnarray}
From the above solution, we observed that stress components ${(\tau_{22})}_1$ and ${(\tau_{33})}_1$ increase with time and tends to $\tau_{\infty}(0)(1+kt)\cos\theta$ and $\tau_{\infty}(0)(1+kt)\sin\theta$ respectively as $t\to\infty$, while the other stress components tend to zero as $t\to\infty$. It is presumed that the geological circumstances and the characteristics of the fault are such that when the stress component ${(\tau_{22})}_1$ attains a critical threshold value $\tau_c<\tau_{\infty}(0)(1+kt)\cos\theta$, the fault starts to move. Studies show that the critical stress lies between $150\times 10^5 N/m^{2}$ to $300 \times 10^5 N/m^{2}$ (Sen et al. 1993). Here, we have considered the critical value $\tau_c$ to be $150$ bar, then the time taken to exceed this critical value is $160.5$ years approx. for $\theta=\frac{\pi}{6}$ and $\alpha=0.5$ which is derived from the expression ${(\tau_{22})}_1$ taking parametric values from table \ref{Table1}.
\subsection{Solutions after the commencement of fault movement}\label{sec4}
In this model, we have assumed a creeping movement when the accumulated stress exceeds a threshold value $\tau_c$ at time $T$. After time $T$, the medium experiences seismic activity for a few seconds. These seismic disturbances gradually diminish, leading to the re-establishment of a quasi-static, aseismic state throughout the medium. We study our model at the time immediately after the restoration of a new aseismic state in the medium, neglecting the seismic period, which usually lasts for a few seconds or a minute at most. Let after time $t=T$, aseismic period occurs again and let $t_1=t-T$ is the new time variable, then for $t_1\geq0$, i.e., for $t\geq T$, the initial values ${(u_i)}_0, {(\tau_{ij})}_0,{(e_{ij})}_0, (i, j=1, 2, 3)$ will be zero. The displacement, stress and strain components after the movement of the fault are ${(u_i)}_2$, ${(\tau_{ij})}_2$ and ${(e_{ij})}_2$ respectively. These components will satisfy constitutive equations (\ref{equation2}), stress equations of motion (\ref{equation3}) and boundary conditions (\ref{equation4})-(\ref{equation9}) for $t_1\geq0$ except for the boundary conditions (\ref{equation6}) and (\ref{equation7}).\\
The modified boundary conditions after the movement of the fault are
\begin{eqnarray}
     {(\tau_{22})}_2\to0 \mbox{ as } |y_2|\to \infty, y_3\geq 0, t_1\geq 0
     \label{equation29}
\end{eqnarray}
\begin{eqnarray}
     {(\tau_{33})}_2\to0 \mbox{ as } y_3\to \infty, -\infty<y_2< \infty, t_1\geq 0
     \label{equation30}
\end{eqnarray} 
Taking the Laplace transformation to the constitutive equations (\ref{equation2}) for the new time coordinate $t_1\geq 0$ and substituting  ${(u_i)}_0= {(\tau_{ij})}_0={(e_{ij})}_0=0$ we get 
\begin{eqnarray}  {(\overline{\tau_{ij}})}_2=\frac{q_1s^{\alpha}+q_2s^{2\alpha}}{1+p_1s^{\alpha}+p_2s^{2\alpha}}\left(\frac{\partial \overline{({u_i})}_2}{\partial y_j}+\frac{\partial \overline{{(u_j})}_2}{\partial y_i}\right);\, i, j=1, 2, 3
\label{equation31}
\end{eqnarray}
where $s$ is the Laplace transform variable concerning $t_1$.\\
Also, after the movement of the fault, it adheres to an additional dislocation condition that defines its creeping motion, which is
\begin{eqnarray}
    {[u_3]}_{F}=U(t_1)f(y_1, y_3)H(t_1)
    \label{equation32}
\end{eqnarray}
where $U(t_1)$ is the slip magnitude, $H(t_1)$ is the Heaviside function defined by
\begin{eqnarray}
H(t_1)=\begin{cases}
    1, t_1> 0\\
    0, t_1\leq 0
\end{cases}
\label{equation33}
\end{eqnarray}
$f(y_1, y_3)$ is continuous slip function of $y_1, y_3$ giving relative displacement across $F$ on depth and length along the fault $F$. This slip function must satisfy the conditions discussed in Sen et al. (1993) for the boundedness of stress and strain components.\\
The discontinuity ${[u_3]}_{F}$ is defined by
\begin{eqnarray}
    {[u_3]}_{F}=\lim_{y_2'\to0^+}u_3-\lim_{y_2'\to0^-}u_3, -L\leq y_1'\leq L, y_2'=0, 0\leq y_3'\leq D
    \label{equation34}
\end{eqnarray}
Let $V$ be the creep velocity of the fault, and then $U(t_1)=Vt_1$. Then for $t_1\geq 0$, (\ref{equation32}) can be written as 
\begin{eqnarray}
    {[u_3]}_{F}=Vt_1f(y_1, y_3)
    \label{equation35}
\end{eqnarray}
Taking Laplace transformation of (\ref{equation35}) we get
\begin{eqnarray}{[\overline{u_1}]}_{F}=\frac{V}{s^2}f(y_1, y_3)
    \label{equation36}
    \end{eqnarray}
Now, since the fault movement follows a dip-slip pattern, displacement along $y_1$ and $y_2$ direction after the fault's movement is negligible, i.e.,
\begin{eqnarray}
    {(u_1)}_2={(u_2)}_2=0
    \label{equation37}
\end{eqnarray}
and the displacement component ${(u_3)}_2$ is calculated by using Green's function technique developed by Maruyama (1964, 1966).\\
Let us take two points $P(\xi_1, \xi_2, \xi_3)$ and $Q(y_1, y_2, y_3)$ such that P is a point on the fault $F$ and Q is any point in the half-space.
Then, using Green's function technique
\begin{eqnarray}
    {(\overline{u_3})}_2=\int\int_{F}{[\overline{u_1}]}_{F}G(P, Q)d\xi_3d\xi_1
    \label{equation38}
\end{eqnarray}
where $G(P, Q)=\frac{\partial}{\partial \xi_2}G_1(P, Q)$ and 
\begin{eqnarray}
    G_1(P, Q)=\frac{1}{\sqrt{(y_1-\xi_1)^2+(y_2-\xi_2)^2+(y_3-\xi_3)^2}}\cr-\frac{1}{\sqrt{(y_1+\xi_1)^2+(y_2-\xi_2)^2+(y_3-\xi_3)^2}}
    \label{equation39}
\end{eqnarray}
Hence 
\begin{eqnarray}
     G(P, Q) = \frac{y_2-\xi_2}{\left\{(y_1-\xi_1)^2+(y_2-\xi_2)^2+(y_3-\xi_3)^2\right\}^{3/2}}\cr-\frac{y_2-\xi_2}{\left\{(y_1+\xi_1)^2+(y_2-\xi_2)^2+(y_3-\xi_3)^2\right\}^{3/2}}
     \label{equation40}
\end{eqnarray}
The coordinate system $(\xi_1, \xi_2, \xi_3)$ is transformed to $(\xi_1', \xi_2', \xi_3')$ following the relation
\begin{eqnarray}
\begin{cases}
        \xi_1=\xi_1'\\ \xi_2=\xi_2'\sin\theta+\xi_3'\cos\theta\\ \xi_3=\xi_2'\cos\theta+\xi_3'\sin\theta+d
\end{cases}
\label{equation41}
\end{eqnarray}
such that on the fault $F$, $\xi_2'=0$.\\
Substituting the expression (\ref{equation40}) in (\ref{equation38}) and using (\ref{equation36}) and (\ref{equation41}) we get,
\begin{align}
    {(\overline{u_3})}_2=\int_{-{L}}^{L}\int_0^{D}\frac{V}{s^2}f(\xi_1', \xi_3')\Big[\frac{y_2-\xi_3'\cos\theta}{\{(y_1-\xi'_1)^2+(y_2-\xi_3'\cos\theta)^2+(y_3-\xi_3'\sin\theta-d)^2\}^{3/2}}\cr-\frac{y_2-\xi_3'\cos\theta}{\{(y_1+\xi'_1)^2+(y_2-\xi_3'\cos\theta)^2+(y_3-\xi_3'\sin\theta-d)^2\}^{3/2}}\Big]\sin\theta d\xi'_1d\xi_3'
    \label{equation42}
\end{align}
which can be written as 
\begin{eqnarray}
    {(\overline{u_3})}_2=\frac{V}{s^2}\Psi_1(y_1, y_2, y_3)
    \label{equation43}
\end{eqnarray}
where
\begin{align}
    \Psi_1=\int_{-{L}}^{L}\int_0^{D}f(\xi_1', \xi_3')\Big[\frac{y_2-\xi_3'\cos\theta}{\{(y_1-\xi'_1)^2+(y_2-\xi_3'\cos\theta)^2+(y_3-\xi_3'\sin\theta-d)^2\}^{3/2}}\cr-\frac{y_2-\xi_3'\cos\theta}{\{(y_1+\xi'_1)^2+(y_2-\xi_3'\cos\theta)^2+(y_3-\xi_3'\sin\theta-d)^2\}^{3/2}}\Big]\sin\theta d\xi'_1d\xi_3'
    \label{equation44}
\end{align}
Taking inverse Laplace transformation of (\ref{equation43}), we get
 \begin{eqnarray}
     {(u_3)}_2=Vt_1H(t_1)\Psi_1(y_1, y_2, y_3)
     \label{equation45}
\end{eqnarray}
If we substitute the expression of ${(\overline{u_i})}_2$, $i=1, 2, 3$ from (\ref{equation37}) and (\ref{equation43}) in the equations (\ref{equation31}) and take inverse Laplace transformation, the stress components after the movement of the faults can be obtained as follows
\begin{eqnarray}
    \begin{cases}
        {(\tau_{11})}_2=0,\, {(\tau_{12})}_2=0,\, {(\tau_{22})}_2=0\\{(\tau_{13})}_2=V\Psi_2H(t_1)\Big[q_1+\frac{q_1p_2-p_1q_2}{p_2\sqrt{p_1^2-4p_2}}\left\{\frac{E_{\alpha}(-\lambda_1t_1^{\alpha})}{-\lambda_1}+\frac{E_{\alpha}(-\lambda_2t_1^{\alpha})}{\lambda_2}\right\}\cr\hspace{9mm}-\frac{q_2}{p_2\sqrt{p_1^2-4p_2}}\left\{\frac{E_{\alpha}(-\lambda_1t_1^{\alpha})}{\lambda_1^2}-\frac{E_{\alpha}(-\lambda_2t_1^{\alpha})}{\lambda_2^2}\right\}\Big]\\{(\tau_{23})}_2=V\Psi_3H(t_1)\Big[q_1+\frac{q_1p_2-p_1q_2}{p_2\sqrt{p_1^2-4p_2}}\left\{\frac{E_{\alpha}(-\lambda_1t_1^{\alpha})}{-\lambda_1}+\frac{E_{\alpha}(-\lambda_2t_1^{\alpha})}{\lambda_2}\right\}\cr\hspace{9mm}-\frac{q_2}{p_2\sqrt{p_1^2-4p_2}}\left\{\frac{E_{\alpha}(-\lambda_1t_1^{\alpha})}{\lambda_1^2}-\frac{E_{\alpha}(-\lambda_2t_1^{\alpha})}{\lambda_2^2}\right\}\Big]\\{(\tau_{33})}_2=2V\Psi_4H(t_1)\Big[q_1+\frac{q_1p_2-p_1q_2}{p_2\sqrt{p_1^2-4p_2}}\left\{\frac{E_{\alpha}(-\lambda_1t_1^{\alpha})}{-\lambda_1}+\frac{E_{\alpha}(-\lambda_2t_1^{\alpha})}{\lambda_2}\right\}\cr\hspace{9mm}-\frac{q_2}{p_2\sqrt{p_1^2-4p_2}}\left\{\frac{E_{\alpha}(-\lambda_1t_1^{\alpha})}{\lambda_1^2}-\frac{E_{\alpha}(-\lambda_2t_1^{\alpha})}{\lambda_2^2}\right\}\Big]
        \label{equation46}
    \end{cases}
\end{eqnarray}
and using the formula $e_{ij}=\frac12\left(\frac{\partial u_i}{\partial y_j}+\frac{\partial u_j}{\partial y_i}\right)$, we will get the strain components as
\begin{eqnarray}
    \begin{cases}
    {(e_{11})}_2=0, \, {(e_{12})}_2=0, \, {(e_{22})}_2=0\\{(e_{13})}_2=\frac12Vt_1\Psi_2(y_1, y_2, y_3)\\
        {(e_{23})}_2=\frac12Vt_1\Psi_3(y_1, y_2, y_3)\\{(e_{33})}_2=Vt_1\Psi_4(y_1, y_2, y_3)
    \end{cases}
    \label{equation47}
\end{eqnarray}
where 
\begin{eqnarray*}
    \Psi_2=\int_{-{L}}^{L}\int_0^{D}f(\xi_1', \xi_3')\Big[ \frac{-3(y_1-\xi_1')(y_2-\xi_3'\cos\theta)}{\{(y_1-\xi'_1)^2+(y_2-\xi_3'\cos\theta)^2+(y_3-\xi_3'\sin\theta-d)^2\}^{5/2}}\cr+\frac{3(y_1+\xi_1')(y_2-\xi_3'\cos\theta)}{\{(y_1+\xi'_1)^2+(y_2-\xi_3'\cos\theta)^2+(y_3-\xi_3'\sin\theta-d)^2\}^{5/2}}\Big]\sin\theta d\xi_1'd\xi_3'
\end{eqnarray*}
\begin{eqnarray*}
   \Psi_3=\int_{-{L}}^{L}\int_0^{D}f(\xi_1', \xi_3')\Big[\frac{1}{\{(y_1-\xi'_1)^2+(y_2-\xi_3'\cos\theta)^2+(y_3-\xi_3'\sin\theta-d)^2\}^{3/2}}\cr-\frac{1}{\{(y_1+\xi'_1)^2+(y_2-\xi_3'\cos\theta)^2+(y_3-\xi_3'\sin\theta-d)^2\}^{3/2}}\cr-\frac{3(y_2-\xi_1'\cos\theta)^2}{\{(y_1-\xi'_1)^2+(y_2-\xi_3'\cos\theta)^2+(y_3-\xi_3'\sin\theta-d)^2\}^{5/2}}\cr+\frac{3(y_2-\xi_3'\cos\theta)^2}{\{(y_1+\xi'_1)^2+(y_2-\xi_3'\cos\theta)^2+(y_3-\xi_3'\sin\theta-d)^2\}^{5/2}}\Big]\sin\theta d\xi_1'd\xi_3'
\end{eqnarray*}
\begin{eqnarray*}
    \Psi_4=\int_{-{L}}^{L}\int_0^{D}f(\xi_1', \xi_3')\Big[ \frac{-3(y_3-\xi_3'\sin\theta-d)(y_2-\xi_3'\cos\theta)}{\{(y_1-\xi'_1)^2+(y_2-\xi_3'\cos\theta)^2+(y_3-\xi_3'\sin\theta-d)^2\}^{5/2}}\cr+\frac{3(y_3-\xi_3'\sin\theta-d)(y_2-\xi_3'\cos\theta)}{\{(y_1+\xi'_1)^2+(y_2-\xi_3'\cos\theta)^2+(y_3-\xi_3'\sin\theta-d)^2\}^{5/2}}\Big]\sin\theta d\xi_1'd\xi_3'
\end{eqnarray*}
Hence substituting (\ref{equation26})-(\ref{equation28}), (\ref{equation37}) and (\ref{equation45})-(\ref{equation47}) in (\ref{equation15}) we get the final solution as
\begin{eqnarray}
    \begin{cases}
        u_1={(u_1)}_0+\frac{\tau_Ly_1}{2q_1}\left[\frac{t^{\alpha}}{\Gamma(\alpha+1)}-\frac{q_2}{q_1}\left(1-E_{\alpha}\left(-\frac{q_1}{q_2}t^{\alpha}\right)\right)\right]\\
        u_2={(u_2)}_0+\frac{\tau_{\infty}(0)y_2\cos\theta}{2q_1}\Big[\frac{t^{\alpha+1}}{\Gamma(\alpha+2)}+\frac{t^{\alpha}}{\Gamma(\alpha+1)}+kt\left(p_1-\frac{q_2}{q_1}\right)\cr\hspace{5mm}+\left\{E_{\alpha }\left(-\frac{q_1}{q_2}t^{\alpha}\right)-1\right\}\left\{\frac{q_2}{q_1}+\frac{kq_2}{q_1}\left(p_1-\frac{q_1p_2}{q_2}-\frac{q_2}{q_1}\right)\right\}\Big]\\
        u_3={(u_3)}_0+\frac{\tau_{\infty}(0)y_3\sin\theta}{2q_1}\Big[\frac{t^{\alpha+1}}{\Gamma(\alpha+2)}+\frac{t^{\alpha}}{\Gamma(\alpha+1)}+kt\left(p_1-\frac{q_2}{q_1}\right)\cr\hspace{5mm}+\left\{E_{\alpha }\left(-\frac{q_1}{q_2}t^{\alpha}\right)-1\right\}\left\{\frac{q_2}{q_1}+\frac{kq_2}{q_1}\left(p_1-\frac{q_1p_2}{q_2}-\frac{q_2}{q_1}\right)\right\}\Big]+Vt_1H(t_1)\Psi_1
    \end{cases}
    \label{equation48}
\end{eqnarray}
\begin{eqnarray}
    \begin{cases}
       \tau_{11}={(\tau_{11})}_0+\left[1+\frac{1}{\sqrt{p_1^2-4p_2}}\left\{\frac{E_{\alpha}\left(-\lambda_1t^{\alpha}\right)}{-\lambda_1}+\frac{E_{\alpha}\left(-\lambda_2t^{\alpha}\right)}{\lambda_2}\right\}\right]\left\{\tau_L-{(\tau_{11})}_0\right\}\\
       \tau_{12}=\frac{{(\tau_{12})}_0}{\sqrt{p_1^2-4p_2}}\left[\frac{E_{\alpha}\left(-\lambda_1t^{\alpha}\right)}{\lambda_1}-\frac{E_{\alpha}\left(-\lambda_2t^{\alpha}\right)}{\lambda_2}\right]\\
         \tau_{13}=\frac{{(\tau_{13})}_0}{{\sqrt{p_1^2-4p_2}}}\left[\frac{E_{\alpha}\left(-\lambda_1t^{\alpha}\right)}{\lambda_1}-\frac{E_{\alpha}\left(-\lambda_2t^{\alpha}\right)}{\lambda_2}\right]\cr\hspace{9mm}+ V\Psi_2H(t_1)\Big[q_1+\frac{q_1p_2-p_1q_2}{p_2\sqrt{p_1^2-4p_2}}\left\{\frac{E_{\alpha}(-\lambda_1t_1^{\alpha})}{-\lambda_1}+\frac{E_{\alpha}(-\lambda_2t_1^{\alpha})}{\lambda_2}\right\}\cr\hspace{9mm}-\frac{q_2}{p_2\sqrt{p_1^2-4p_2}}\left\{\frac{E_{\alpha}(-\lambda_1t_1^{\alpha})}{\lambda_1^2}-\frac{E_{\alpha}(-\lambda_2t_1^{\alpha})}{\lambda_2^2}\right\}\Big]\\
         \tau_{22}=\tau_{\infty}(0)(1+kt)\cos\theta+\frac{1}{\sqrt{p_1^2-4p_2}}\left[\frac{E_{\alpha}\left(-\lambda_1t^{\alpha}\right)}{\lambda_1}-\frac{E_{\alpha}\left(-\lambda_2t^{\alpha}\right)}{\lambda_2}\right]\cr\hspace{9mm}\left\{{(\tau_{22})}_0-{\tau_{\infty}}(0)\cos\theta\right\}\\
         \tau_{23}=\frac{{(\tau_{23})}_0}{{\sqrt{p_1^2-4p_2}}}\left[\frac{E_{\alpha}\left(-\lambda_1t^{\alpha}\right)}{\lambda_1}-\frac{E_{\alpha}\left(-\lambda_2t^{\alpha}\right)}{\lambda_2}\right]\cr\hspace{9mm}+ V\Psi_3H(t_1)\Big[q_1+\frac{q_1p_2-p_1q_2}{p_2\sqrt{p_1^2-4p_2}}\left\{\frac{E_{\alpha}(-\lambda_1t_1^{\alpha})}{-\lambda_1}+\frac{E_{\alpha}(-\lambda_2t_1^{\alpha})}{\lambda_2}\right\}\cr\hspace{9mm}-\frac{q_2}{p_2\sqrt{p_1^2-4p_2}}\left\{\frac{E_{\alpha}(-\lambda_1t_1^{\alpha})}{\lambda_1^2}-\frac{E_{\alpha}(-\lambda_2t_1^{\alpha})}{\lambda_2^2}\right\}\Big]\\
          \tau_{33}=\tau_{\infty}(0)(1+kt)\sin\theta+\frac{1}{\sqrt{p_1^2-4p_2}}\left[\frac{E_{\alpha}\left(-\lambda_1t^{\alpha}\right)}{\lambda_1}-\frac{E_{\alpha}\left(-\lambda_2t^{\alpha}\right)}{\lambda_2}\right]\cr\hspace{9mm}\left\{{(\tau_{33})}_0-{\tau_{\infty}}(0)\sin\theta\right\}+2V\Psi_4H(t_1)\Big[q_1+\frac{q_1p_2-p_1q_2}{p_2\sqrt{p_1^2-4p_2}}\cr\hspace{9mm}\left\{\frac{E_{\alpha}(-\lambda_1t_1^{\alpha})}{-\lambda_1}+\frac{E_{\alpha}(-\lambda_2t_1^{\alpha})}{\lambda_2}\right\}-\frac{q_2}{p_2\sqrt{p_1^2-4p_2}}\left\{\frac{E_{\alpha}(-\lambda_1t_1^{\alpha})}{\lambda_1^2}-\frac{E_{\alpha}(-\lambda_2t_1^{\alpha})}{\lambda_2^2}\right\}\Big] 
    \end{cases}
    \label{equation49}
\end{eqnarray}
\begin{eqnarray}
    \begin{cases}
        e_{11}={(e_{11})}_0+\frac{\tau_L}{2q_1}\left[\frac{t^{\alpha}}{\Gamma(\alpha+1)}-\frac{q_2}{q_1}\left(1-E_{\alpha}\left(-\frac{q_1}{q_2}t^{\alpha}\right)\right)\right]\\
        e_{12}={(e_{12})}_0\\
        e_{13}={(e_{13})}_0+\frac12Vt_1\Psi_2(y_1, y_2, y_3)\\
        e_{22}={(e_{22})}_0+\frac{\tau_{\infty}(0)\cos\theta}{2q_1}\Bigg[\frac{t^{\alpha+1}}{\Gamma(\alpha+2)}+\frac{t^{\alpha}}{\Gamma(\alpha+1)}+kt\left(p_1-\frac{q_2}{q_1}\right)\cr\hspace{5mm}+\left\{E_{\alpha }\left(-\frac{q_1}{q_2}t^{\alpha}\right)-1\right\}\left\{\frac{q_2}{q_1}+\frac{kq_2}{q_1}\left(p_1-\frac{q_1p_2}{q_2}-\frac{q_2}{q_1}\right)\right\}\Bigg]\\
        e_{23}={(e_{23})}_0+\frac12Vt_1\Psi_3(y_1, y_2, y_3)\\
        e_{33}={(e_{33})}_0+\frac{\tau_{\infty}(0)\sin\theta}{2q_1}\Bigg[\frac{t^{\alpha+1}}{\Gamma(\alpha+2)}+\frac{t^{\alpha}}{\Gamma(\alpha+1)}+kt\left(p_1-\frac{q_2}{q_1}\right)\cr\hspace{5mm}+\left\{E_{\alpha }\left(-\frac{q_1}{q_2}t^{\alpha}\right)-1\right\}\left\{\frac{q_2}{q_1}+\frac{kq_2}{q_1}\left(p_1-\frac{q_1p_2}{q_2}-\frac{q_2}{q_1}\right)\right\}\Bigg]+Vt_1\Psi_4(y_1, y_2, y_3)   
    \end{cases}
    \label{equation50}
\end{eqnarray}
\section{Numerical Computation}\label{sec4}
To study the nature of the surface displacement, surface shear stress and strain, we assigned values of the model parameters by following the articles Cathles (1975), Clift (2002), Aki (1980), Karato (2010), Kundu et al. (2021). These parameters are chosen on the basis of research works done by various scholars using field data from real-world physical phenomena. All the parametric values with their symbol have been given in table \ref{Table1}.
\begin{table}[ht!]
\centering
\begin{tabular}{ |c|c| } 
 \hline
 Parametric symbol & Parametric values\\ 
 \hline
 Rigidity ($\mu_1$) & $3.5\times 10^{10} N/m^2$\\ 
 Rigidity ($\mu_2$) & $3\times10^{10} N/m^2$\\ 
 Viscosity ($\eta_1$) & $3.5\times 10^{19}$  Pascals\\
 Viscosity ($\eta_2$) & $3\times 10^{19}$  Pascals\\
 Initial stresses $ \left({(\tau_{ij})}_0\right)$ & $50\times 10^5 N/m^2$\\
 Initial stress far away from the fault $\left({\tau_{\infty}}(0)\right)$ & $20\times 10^5 N/m^2$\\
 Length of the fault $(2L)$ & 20,000 m\\
 Width of the fault $(D)$ & $10,000$ m\\
 Depth of the fault $(d)$ & $1000$ m\\
 Constant ($k$) & $10^{-9}$\\
 Angle of inclination ($\theta$) & $\frac{\pi}{6}$, $\frac{\pi}{4}$, $\frac{\pi}{3}$, $\frac{\pi}{2}$\\
 Velocity $(V)$ & $0.01, 0.03, 0.05$ meter/year\\
 Order of the fractional derivative $(\alpha)$ & $0.1, 0.4, 0.5, 0.7, 1$\\
 Time $(T)$ &$160.5$ years\\
 \hline
 \end{tabular}
 \vspace{4mm}
\caption{Parametric values with their symbols used for numerical computation}
    \label{Table1}
\end{table}\\
The following slip function has been taken to represent the crack of the surface after the fault's movement (Kundu et al. 2021).\\
$f(y_1', y_3')=R\left(1-\frac{1}{L^2{y_1'}^2}\right)\left(1-\frac{3{y_3'}^2}{D^2}+\frac{3{y_3'}^3}{D^2}\right)$ 
where $R=1$ cm is the magnifying factor.\\

\begin{figure}[ht!]
\centering
\includegraphics[width=0.7\textwidth]{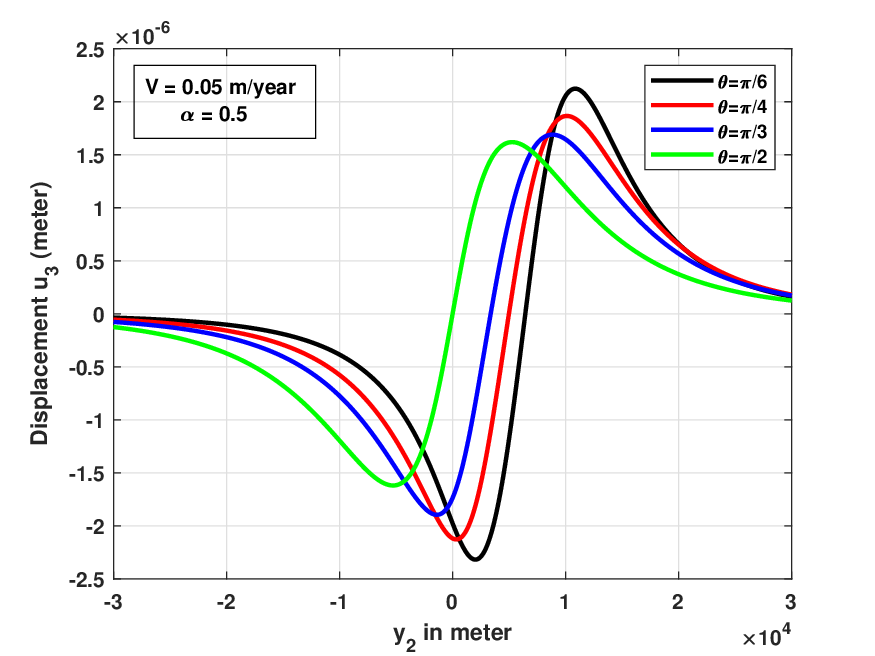}
    \caption{Change in surface displacement $u_3$ against $y_2$ for different inclination.}
    \label{figure2} 
\end{figure}

The surface displacement, stress and strain components were analyzed with the help of the parametric values mentioned in table \ref{Table1}, with the new time origin $t_1=t-T=1$ year where $T=160.5$ years. MATLAB R2022b has been used to plot the following figures.\\\\
\textbf{Change of displacement:}\\
The surface displacement
$u_3= {(u_3)}_0+\frac{\tau_{\infty}(0)y_3\sin\theta}{2q_1}\Big[\frac{t^{\alpha+1}}{\Gamma(\alpha+2)}+\frac{t^{\alpha}}{\Gamma(\alpha+1)}+kt\left(p_1-\frac{q_2}{q_1}\right)+\left\{E_{\alpha }\left(-\frac{q_1}{q_2}t^{\alpha}\right)-1\right\}\left\{\frac{q_2}{q_1}+\frac{kq_2}{q_1}\left(p_1-\frac{q_1p_2}{q_2}-\frac{q_2}{q_1}\right)\right\}\Big]+Vt_1H(t_1)\Psi_1(y_1, y_2, y_3)$\\\\
We measure the surface displacement against $y_2$ for different inclinations $\theta$. Here, creep velocity is taken as $V=0.05$  m/year and order of fractional derivative $\alpha=0.5$. From figure \ref{figure2}, it has been observed that the surface displacement significantly depends on the inclination of the fault. Although there are some notable differences depending upon the inclinations of the fault, there are some common characteristics as well, irrespective of the inclination $\theta$. (i) The maximum magnitude of surface displacement attains close to the fault for both $y_2>0$ and $y_2<0$. (ii) As $y_2$ increases ($|y_2|>>D, D=10000$ m), the magnitude of displacement diminishes rapidly (this rate of decreasing is higher for smaller values of $\theta$) and displacement tends to zero as $|y_2|\to\infty$.

\begin{figure}[ht!]
\centering
\includegraphics[width=0.7\textwidth]{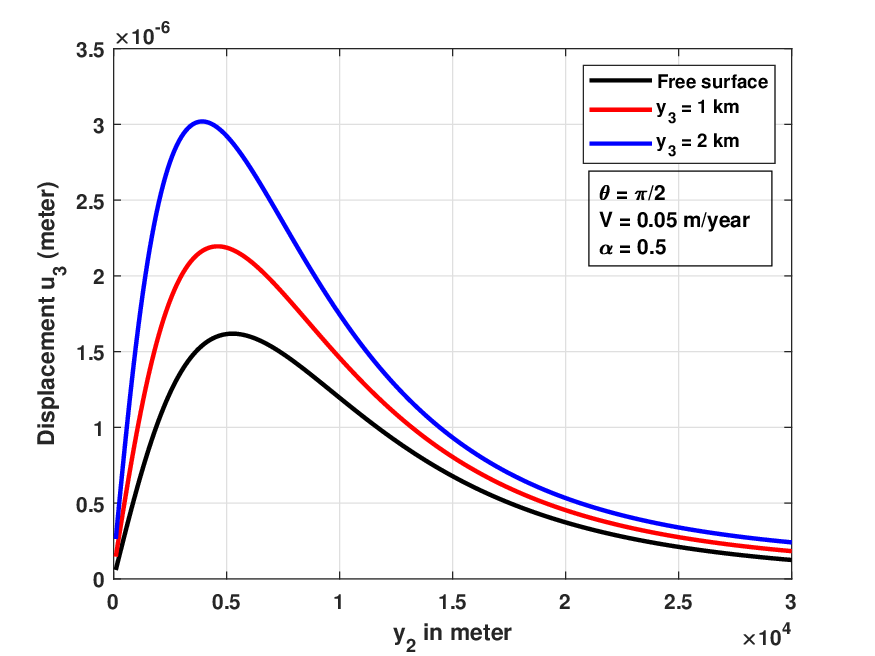}
    \caption{Displacement $u_3$ along $y_2$ for different depth due to dislocation.}
    \label{figure3}
\end{figure}

\begin{figure}[ht!]
\centering
\includegraphics[width=0.7\textwidth]{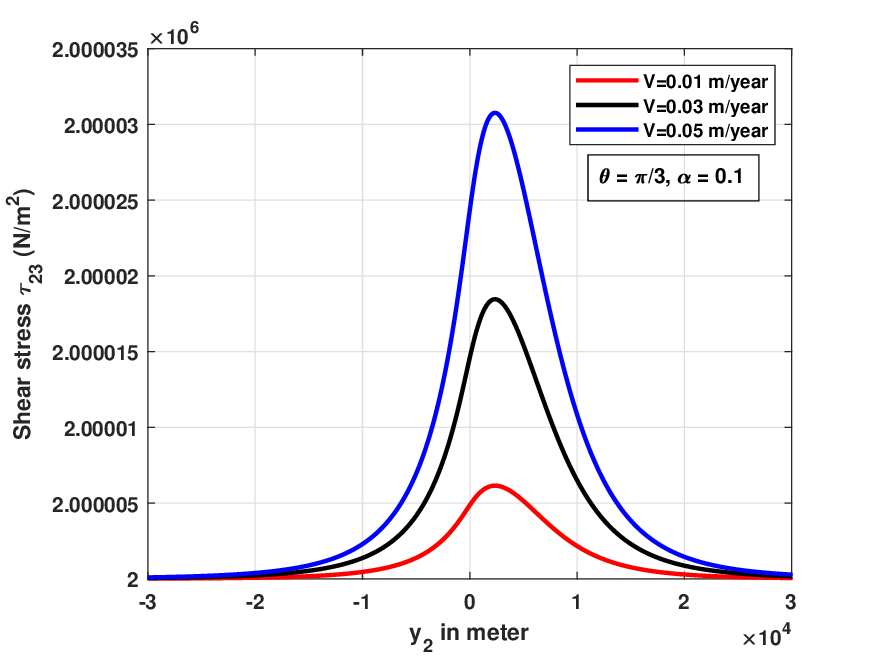}
    \caption{Change in surface shear stress $\tau_{23}$ against $y_2$ for different creeping velocities.}
    \label{figure4}
\end{figure}

\begin{figure}[ht!]
\centering
\includegraphics[width=0.7\textwidth]{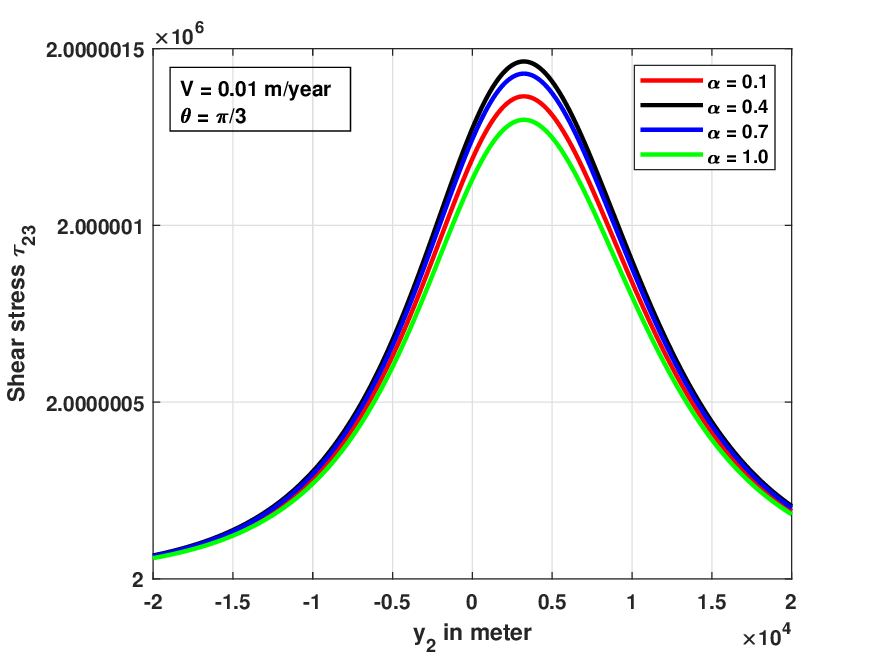}
    \caption{Change in surface shear stress $\tau_{23}$ against $y_2$ for different fractional order $\alpha$.}
    \label{figure5}
\end{figure}

\begin{figure}[ht!]
\centering
\includegraphics[width=0.7\textwidth]{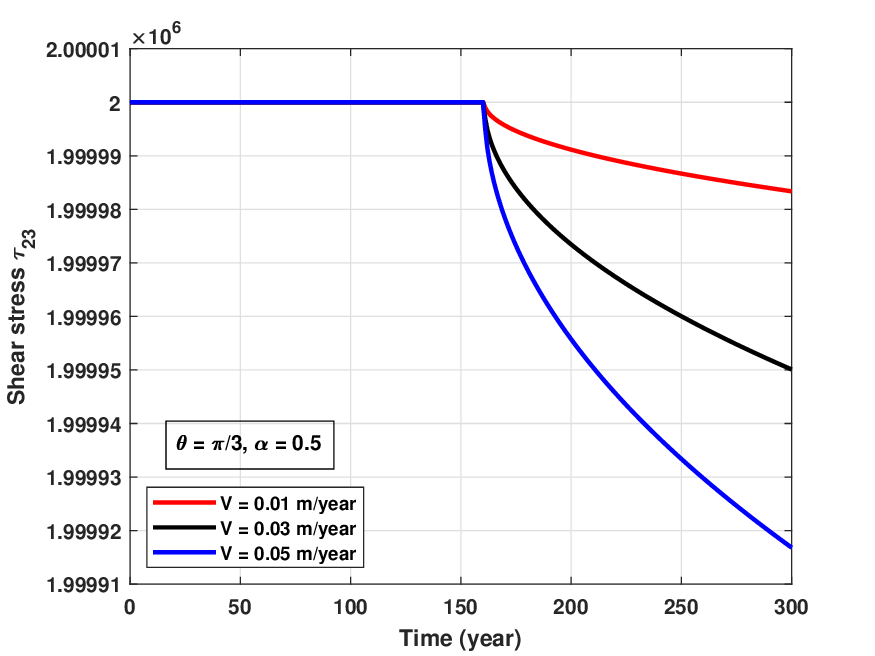}
    \caption{Change in surface shear stress $\tau_{23}$ against time $t$ for different fractional order $\alpha$.}
    \label{figure6}
\end{figure}

\begin{figure}[ht!]
\centering
\includegraphics[width=0.7\textwidth]{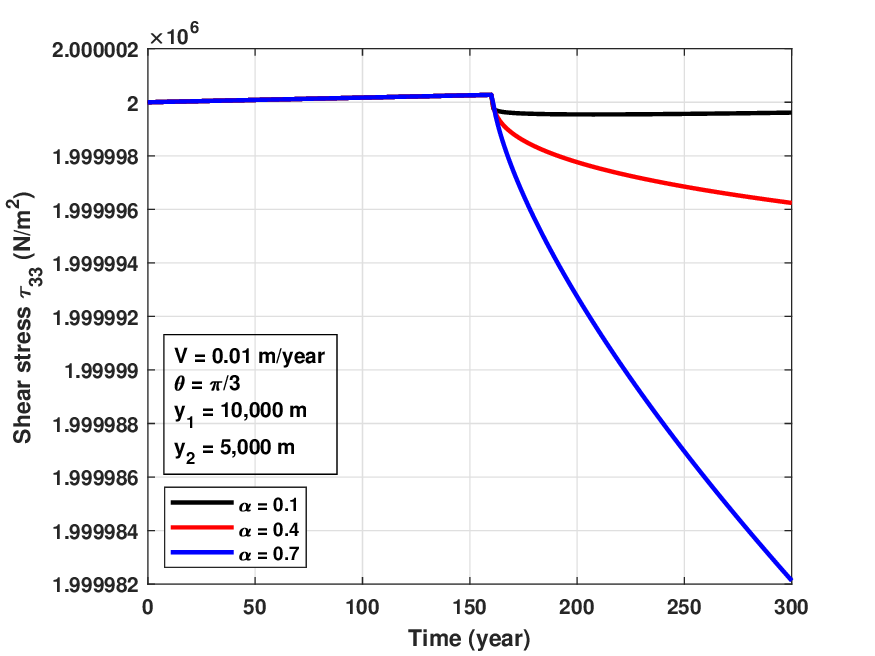}
    \caption{Change in normal stress $\tau_{33}$ against time $t$ for different fractional order $\alpha$.}
    \label{figure7}
\end{figure}

\begin{figure}[ht!]
\centering
\includegraphics[width=0.7\textwidth]{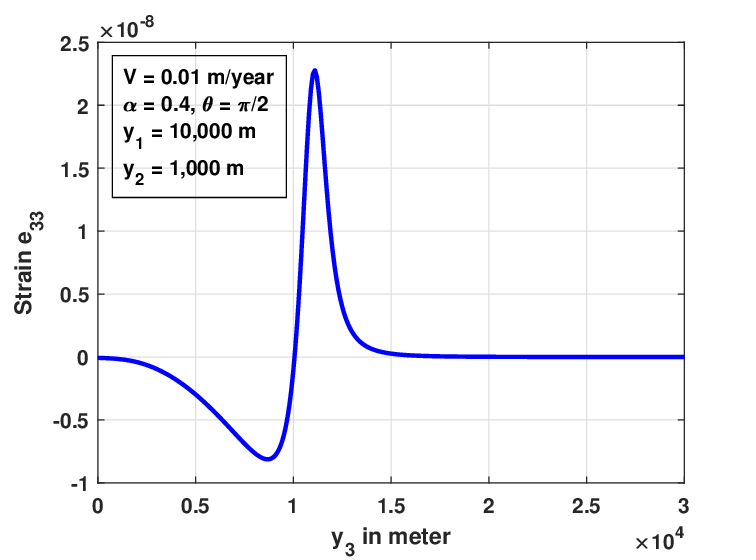}
    \caption{Changes of strain $e_{33}$ against $y_3$ for when $\theta=\frac{\pi}{2}, V=0.01$ m/year, $\alpha=0.4$.}
    \label{figure8}
\end{figure}

On the right side of the fault, the peak of surface displacement for $\theta=\frac{\pi}{6}$ is approx. $2.1211\times 10^{-6}$ m/year ($\approx 0.0021$ mm/year) while that for $\theta=\frac{\pi}{2}$ is approx. $1.6193\times 10^{-6}$ m/year ($\approx 0.00162$ mm/year), which supports the observation of Nicol et al. (1997).
\\\\
Figure \ref{figure3} illustrates the displacement component $u_3$ along $y_2$ for different depths below the free surface due to the fault movement. In this case, creep velocity is $V=0.05$ m/year, $\alpha=0.5$, and the fault is taken to be vertical. It has been observed that the rate of change in displacement accelerates with depth and follows a consistent pattern at all depths. The maximum magnitude is higher for $y_3=2$ km and is comparatively lower on the free surface, i.e., as we move vertically downwards towards the fault, the magnitude of displacement is higher. For any $y_3$, displacement tends to zero as $y_2\to\infty$, which aligns with geophysical findings.
\\
\textit{\textbf{Change of stress:}}\\
Next, we consider the variation of surface shear stress $\tau_{23}$ against $y_2$ for different creep velocities. Here, fractional order and angle of inclination are taken as$\alpha=0.1$ and $\theta=\frac{\pi}{3}$, respectively. It is observed from the figure \ref{figure4} that surface shear stress $\tau_{23}$ starts to accumulate after the commencement of fault creep in the order of $10^6 N/m^2$. The maximum magnitude of surface shear stress $\tau_{23}$ attains at a distance $y_2=2500$ meter (approx.) for each $V$. The maximum magnitude of $\tau_{23}$ for $V=0.05$ m/year is approx. $2.00003076\times 10^6 N/m^2$ and that is for $V=0.01$ m/year is approx. $2.0000061\times 10^6 N/m^2$. As the distance from the fault increases, the surface stress diminishes rapidly at an accelerated rate, and this effect becomes less pronounced as $|y_2|$ increases.
\\\\
The impact of fractional order $\alpha$ on surface shear stress $\tau_{23}$ against $y_2$ is illustrated in figure \ref{figure5}, taking creep velocity $V=0.01$ m/year and inclination $\theta=\frac{\pi}{3}$. All curves exhibit a similar trend, with shear stress $\tau_{23}$ peaking near the fault ($y_2\approx 3400$ meter) and decreasing nearly symmetrically as we move away from the fault. The peak values of shear stress are slightly different for each $\alpha$. It is maximum for $\alpha=0.4$ and is minimum for $\alpha=1$. The difference in magnitude of shear stress is not so prominent for each $\alpha$.
\\
Figure \ref{figure6} shows the shear stress accumulation with time for different creep velocities. It is observed that the creep velocity $V$ is significantly influenced after dislocation. The higher value of $V$ leads to a more rapid decrease in shear stress. The red curve (for $V=0.01$ m/year) shows a slight decrease in shear stress reaching around $1.999983\times10^6$ at $300$ years while the blue curve (for $V=0.05$ m/year) shows a comparatively significant decrease of shear stress reaching around $1.999913\times10^6$ by that time. i.e., the rate of this accumulation of surface shear stress is directly proportional to the value of creep velocities $V$.
\\
Figure \ref{figure7} displays the plot of the normal stress $\tau_{33}$ against time $t$ for different orders of the fractional derivative. For each $\alpha$, $\tau_{33}$ increases very slowly till approx. $160.5$ years. After exceeding the critical stress in the year $160.5$, this surface shear stress $\tau_{33}$ decreases slowly with different rates. The black curve ($\alpha=0.1$) indicates a very gradual decrease over time, while the blue curve ($\alpha=0.7$) represents the most significant decrease in that interval of time.
\\\\
\textit{\textbf{Change of strain:}}\\
We examine the variation of strain $e_{33}$ against $y_3$ on the aseismic period after critical time $T$ through figure \ref{figure8}. The magnitude of change of strain is found to be in the order of $10^{-8}$, which confirms the strain rate value studied by Michael (2005). A significant change in strain is observed around a depth of $10,000$ m below the free surface, where the fault is situated. The strain is maximum at $y_3=11,100$ m, which is approximately $2.278\times 10^{-8}$ beyond which the strain $e_{33}$ gradually returns towards zero. Such behaviour could indicate a strain concentration or a localized deformation zone within the material.
\section{Conclusion \& Future Scope}\label{sec5}
In this paper, we derive explicit expressions for the displacement, stress, and strain components of a finite, buried, inclined dip-slip fault within a viscoelastic half-space characterized by Burger's rheology. Using this model, we have focused on analyzing changes in surface displacement, the pattern of stress accumulation/ release and variations in strain for different fault inclinations, creep velocities, and fractional orders. Results show that the changes of displacement component $u_3$ are significantly influenced by the inclination of the fault, though the pattern of change of this displacement is notably similar (figure \ref{figure2}). We found that the total displacement of an infinite fault in a classical integer Burger's medium is smaller than that of a finite fault in the fractional Burger's medium, as compared to the findings of (Kundu et al. 2021). This observation aligns with real-world examples (Nicol et. al, 1997). 
The changes in surface shear stress components are affected by the velocity of the fault movement and the fractional model. The decrease in shear stress over time could indicate material degradation or stress relaxation in the medium.\\

Our result could provide valuable insights into the stress/strain distribution in the system under study, highlighting regions of interest where significant stress/strain variations occur.\\

The concept presented in this paper can be extended to include interacting faults in future research. Additionally, the analysis could be applied to non-planar fault models as well.
\section*{Declarations}
\begin{itemize}
\item \textbf{Acknowledgement}\\
The first author expresses gratitude to the National Institute of Technology, Durgapur, India, for providing logistical and financial support for this study.
\item \textbf{Conflict of interest}\\
The authors confirm that there are no conflicts of interest.
\item \textbf{Availability of data and materials}\\
We have used data taken from some published papers and their
Citations corresponding to the text have been shown. 
\item \textbf{Code availability}\\
The required code was implemented using MATLAB R2022b software.
\item \textbf{Authors' contributions}\\
The first author was responsible for drafting the original manuscript and played a crucial role in developing and implementing the required code for the study. The second author contributed through supervision, manuscript revisions, and improving the overall linguistic quality of the paper.
\end{itemize}

\end{document}